\documentclass{ws-rv9x6}
\usepackage{subfigure}   
\usepackage{ws-rv-thm}   
\usepackage{ws-rv-van}   
\makeindex
\usepackage{wrapfig}
\usepackage{epsfig}
\begin{document}

\chapter[]{Parton correlations effects in double
parton distribution functions} 

\author[M.Rinaldi]{M. Rinaldi\footnote{Email: mrinaldi@ific.uv.es}}
\address{Instituto de Fisica Corpuscular (CSIC-Universitat de 
Valencia), 
Parc Cientific UV, C/ Catedratico Jose Beltran 2, E-46980 Paterna 
(Valencia), Spain.}

\author[Fede]{Federico Alberto Ceccopieri}
\address{Dipartimento di Fisica e Geologia, Universit\`a degli Studi di Perugia, 
via A. Pascoli, 06123 Perugia, Italy
and
IFPA, Universit\'e de Li\`ege, B4000, Li\`ege, Belgium}

\author[]{Sergio Scopetta}
\address{Dipartimento di Fisica e Geologia, Universit\`a degli Studi di 
Perugia, Via A. Pascoli, 06123, Italy and Istituto Nazionale di Fisica 
Nucleare, Sezione di Perugia, Perugia, Italy}

\author[]{Marco Traini}
\address{INFN - TIFPA, Dipartimento di Fisica, 
Universit\`a degli Studi di Trento, Via Sommarive 14, I-38123 Povo, Trento, 
Italy}

\author[]{Vicente Vento}
\address{ Departament de Fisica Teòrica, Universitat de Val\'encia, Institut 
de Fisica Corpuscular, Consejo Superior de Investigaciones Científicas, 46100 
Burjassot,
València, Spain}


\begin{abstract}
Double parton distribution functions (dPDFs), measurable in hadron-hadron 
collisions and encoding  information on how partons inside a proton are 
correlated among each
other, could represent a new tool to explore the three dimensional partonic 
structure of hadrons. 
In the present contribution, results of the calculations of
dPFDs are  presented.
Phenomenological calculations of experimental observables, 
sensitive to dPDFs  are also discussed showing how double parton 
correlations could be estimated in the next LHC run.
\end{abstract}

\body

\section{Introduction}
Since a long time it has been established that  
 multiple parton interactions (MPI) are a fundamental piece in
 the investigation on hadron-hadron collisions at the LHC. 
In these kind of events, more than one parton of a hadron 
interact with  partons of the other colliding hadron. However, the
 MPI contribution  to the total cross section is  
suppressed  with respect to the single partonic interaction.
Nevertheless,
MPI could represent a background for the search of new  Physic at the 
LHC and 
the measurement of their cross sections  is an 
important experimental
 challenge. From a theoretical point of view,  one of the main 
interests on MPI is
the 
possibility of accessing new fundamental information on the partonic proton 
structure. To this aim,  we focused our attention on the   double parton 
scattering (DPS), the most simple case of MPI, which can be 
observed, in principle, 
 in several processes, e.g., $WW$  with dilepton productions 
and double
Drell-Yan processes (see, Refs. \cite{3a,4a,5a,6a, report} for
recent 
reviews).
At the LHC, DPS has been
observed some years ago \cite{16a} and represents  also a
background for the Higgs production in several channels.
Formally, 
the DPS cross section
depends on  the so called double 
parton distribution functions (dPDFs),
$F_{ij}(x_1,x_2,{\vec z}_\perp,\mu)$, which describe the 
joint probability 
of finding two partons of flavors $i,j=q, \bar q,g$ with 
longitudinal momentum fractions $x_1,x_2$ and distance 
$\vec z_\perp$ in the transverse plane inside the hadron, see Ref. \cite{1a}~.
Here $\mu$ is the renormalization scale.
Since for the moment being no data are available for dPDFs, they 
are 
 usually approximated in experimental analyses through a fully factorized 
ansantz:

\begin{align}
 F_{ij}(x_1,x_2, \vec z_\perp,\mu) = f_i(x_1,\mu)f_j(x_2,\mu) T(\vec z_\perp),
 \label{app}
\end{align}
being $f_i(x,\mu)$ the usual
 one-body parton distribution functions (PDFs) and  $T(\vec z_\perp)$, a 
distribution 
encoding the probability of 
finding two partons at distance $z_\perp$.
The  assumption  (\ref{app}) is to neglects all possible unknown  
 double parton
correlations (DPC) between the two interacting partons  (e.g.  Ref. 
\cite{kase1} ).
Being 
 dPDFs  non
perturbative quantities in QCD,  they can not be easily evaluated 
and
constituent quark model (CQM) calculations   could help to grasp their basic 
features checking if  the factorized approximation is a suitable 
ansatz for dPDFs see Refs. \cite{man,noi1,noij1,bru, 
noice}~. 
Double PDFs
  are first of all calculated at the low
hadronic scale of the model, $\mu_0$, then, in order
 to compare the outcome with future data taken at
high momentum transfer, $Q > \mu_0$, it is  necessary to perform a
perturbative QCD (pQCD) evolution  by 
using  dPDF evolution equations, see Refs. \cite{23a,24a,blok_1}.
In particular, this step is fundamental to understand to what extent  DPC 
survive at the kinematic conditions of experiments.
Thanks to this procedure, future data analyses
of the DPS processes could be guided, in principle, by 
model calculations.
To this aim in Refs. \cite{noij1, noij2}~, DPC in dPDFs have been studied at  
the energy 
scale of the experiments and  the so called  
$\sigma_{eff}$,  evaluated in Refs. \cite{noiPLB,  noice, 
noiads}~. In fact,  DPS cross 
section, in 
processes with final 
state 
$A+B$, is written through   the following ratio  (see 
e.g.~Ref.~\cite{MPI15}):

\begin{eqnarray}
\sigma^{A+B}_{DPS}  = \dfrac{m}{2} \dfrac{\sigma_{SPS}^A 
\sigma_{SPS}^B}{\sigma_{eff}}\,,
\label{sigma_eff_exp}
\end{eqnarray}
where $m$ is combinatorial factor depending on the final states $A$ and $B$ 
($m=1$ for $A=B$ or $m=2$ for $A \neq B$) and
$\sigma^{A(B)}_{SPS}$ is the single parton scattering cross section with final 
state $A(B)$. 
The present knowledge on DPS cross sections
has been condensed in the experimental  extraction of 
$\sigma_{eff}$~\cite{MPI15,S1,S2,S3,S4,S5,S6,S7}. 
A constant value, $\sigma_{eff} \simeq$ 
15 mb, is compatible, within errors, with data: 
a
result  obtained within  the fully uncorrelated ansatz for 
dPDFs   ( all DPC are neglected). 
$\sigma_{eff}$ can be calculated  within CQM in 
order to characterize 
``signals'' of DPC.
In the next sections the 
  possible dependence  of $\sigma_{eff}$ on the 
longitudinal momentum fraction carried by the interacting 
partons is presented,
see Refs.  \cite{noiPLB,  noice, 
noiads}~. Furthermore a phenomenological 
calculation of DPS cross section  for the  $WW$ same sign production, is 
discussed: the main 
outcome of this last investigation  is that DPC could be  accessed in the next 
LHC run.

\section{Calculation of dPDFs in CQM and pQCD evoultion}
In  our first studies,  dPDFs have been calculated within CQM, 
see Refs. \cite{noi1, noij1}~, and it has been shown that at the hadronic  
scale, DPC can not be neglected, in particular the factorization on the $x_1$ 
and $x_2$ dependence is never supported. Furthermore, in Ref. 
\cite{noice}~, it has been demonstrated that also the factorization in the 
$\vec z_\perp-(x_1,x_2)$ dependence  of dPDFs, evaluated within the 
relativistic 
Light-Front (LF) approach \cite{pol,pauli}, does not represent a good 
approximation for dPDFs at the 
hadronic  scale, independently on the chosen model for the proton wave 
function. The pQCD 
evolution of  model calculations of dPDFs has been investigated and addressed 
in Ref. 
\cite{noij1,noij2}~.
A deep 
study on the role of perturbative and non perturbative correlations on 
dPDFs at high energy scales has been presented in Ref. \cite{noij2}~. To this 
aim, the following 
ratios have been defined:

\begin{align}
\label{eq:ratioab2}
 ratio_{ab} &= \dfrac{F_{ab}(x_1,x_2 = 0.01, k_\perp=0; 
Q^2)}{a(x_1;Q^2)b(x_2 = 0.01;Q^2)}~;
\\
\label{eq:ratioabP}
 ratio_{ab}^P &= \dfrac{F_{ab}(x_1,x_2 = 0.01, k_\perp=0; 
Q^2)|^P}{a(x_1;Q^2)b(x_2 = 0.01;Q^2)};\\
\label{eq:ratioabNP}
ratio_{ab}^{NP} &= \dfrac{F_{ab}(x_1,x_2 = 0.01, k_\perp=0; 
Q^2)}{F_{ab}(x_1,x_2 = 0.01, k_\perp=0; 
Q^2)|^P}
\end{align}

where here:

\begin{align}
\label{eq:pdpdf}
 F_{ab}(x_1,x_2 = 0.01, k_\perp=0; 
Q^2)|^P = \big[a(x_1;Q^2)b(x_2 = 0.01;Q^2)  \big]^{dPDF evolution}~.
\end{align}
In the above expressions $a(x;Q^2)$ and $b(x;Q^2)$  are 
the single 
PDFs for two given 
partons of flavor $a$ and $b$ and $Q^2 = 250 $ GeV$^2$.  
The ratio  (\ref{eq:ratioab2}) would be equal to one if
the factorized ansatz, Eq. (\ref{app}) 
holds.
However, since in this quantity  the numerator and  
 the denominator 
evolve  by means of different evolution equations, $ratio_{ab}$ is sensitive to 
non perturbative correlations, encoded in 
the 
proton wave function used to calculate dPDFs and PDFs, and to the perturbative 
ones due to the difference 
in 
the evolution equations of dPDFs and PDFs. 
In order to disentangle these two 
different contributions, the ratios Eqs. (\ref{eq:ratioabP}, 
\ref{eq:ratioabNP}) have been introduced and  calculated. In addition, the 
quantity  Eq. (\ref{eq:pdpdf}), appearing in Eqs. (\ref{eq:ratioabP}, 
\ref{eq:ratioabNP}),  is evaluated by evolving the product of PDFs through the 
dPDF evolution scheme. Thanks to this feature,  $ratio_{ab}^P$ is sensitive 
to perturbative correlations, in fact,
its  numerator and 
denominator   evolve 
differently in pQCD but having the same input 
at the initial scale. On the other hand, 
since in $ratio_{ab}^{NP}$,  the numerator and denominator evolve within the 
same scheme, but using a different  input at the initial scale, this quantity
 is sensitive to non 
perturbative correlations.

\begin{figure}[t]
\vspace{15.0cm}
\includegraphics{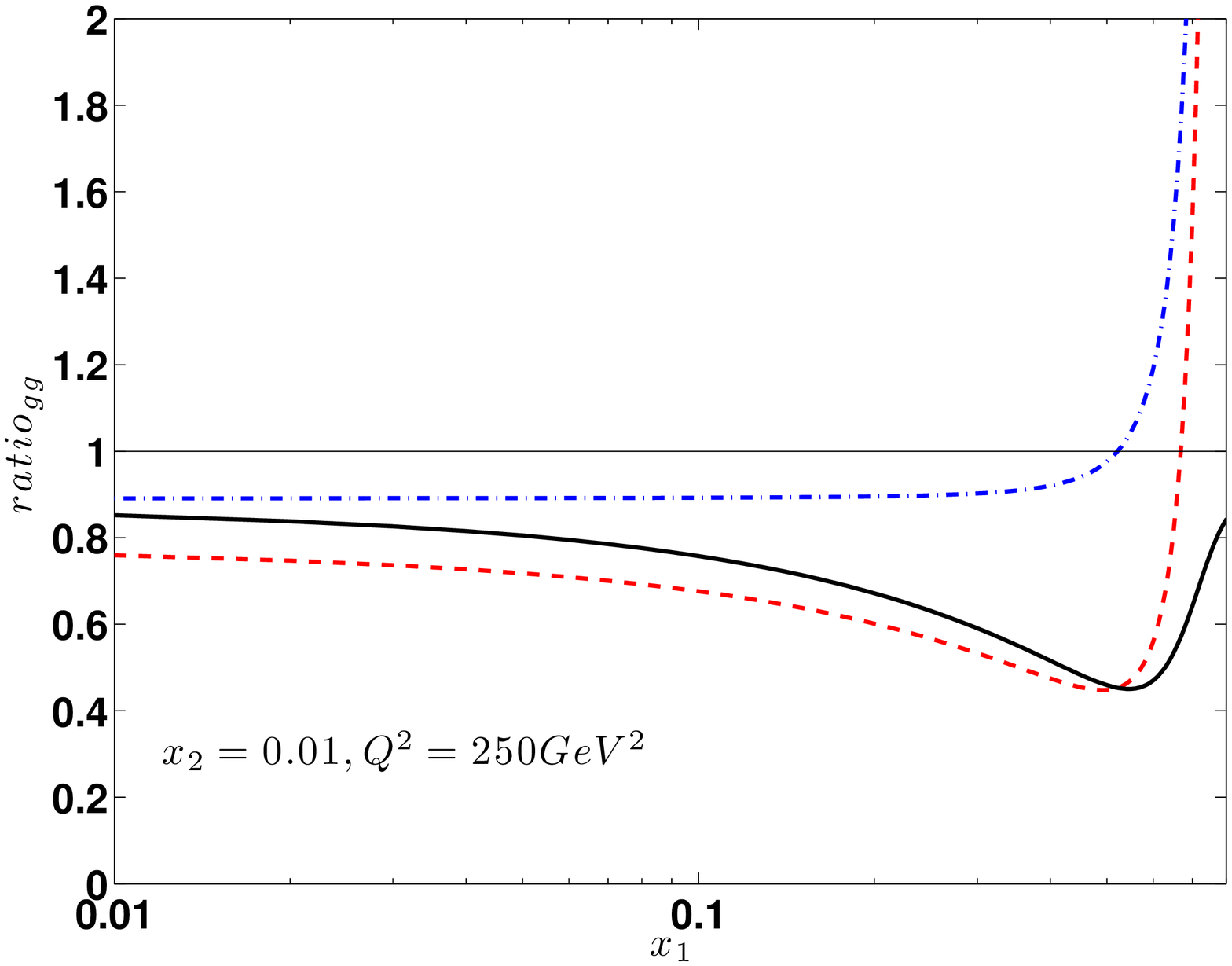}
\includegraphics{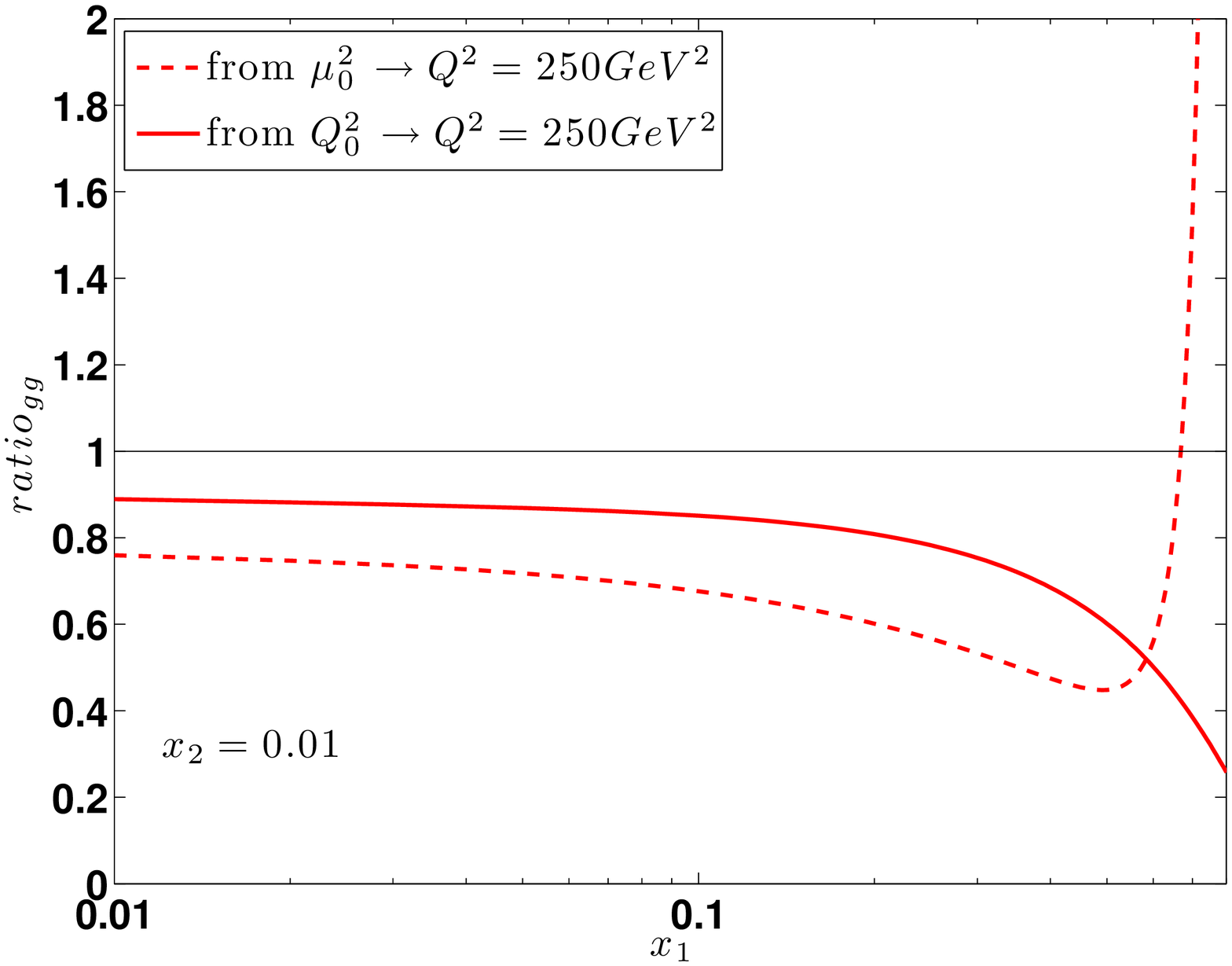}
\vskip -10cm
\caption{ \footnotesize 
Left panel: gluon gluon ratios $ratio_{gg}$ (dashed line), $ratio_{gg}^P$ 
(dot-dashed) and $ratio_{gg}^{NP}$ (continuous line) evaluated at the scale 
$Q^2 = 250$ GeV$^2$ and for $x_2 = 0.01$. Right panel: the same ratio 
evaluated including non perturbative sea quarks and gluons}. 
\label{fig:ratiogg&uVuV_mCS}
\end{figure}
An example  of these ratios is shown  in Fig. 
\ref{fig:ratiogg&uVuV_mCS} (left panel) 
for, e.g.,  $a=b=g$, being gluons the  
highest partonic component at LHC kinematics. In particular, here use has been 
made of the dPDFs discussed in Ref. \cite{noij1} and obtained within the CQM 
described in Ref. \cite{faccioli}. The full ratio $ratio_{ab}$, Eq. 
(\ref{eq:ratioab2}), 
(dashed line),  influenced by both perturbative (dot-dashed line) 
and non-perturbative (continuous line) effects, is compared with those
where perturbative and 
non-perturbative correlations are disentangled.
As one can see in the left panel of Fig.  \ref{fig:ratiogg&uVuV_mCS},  for the 
gluon-gluon distribution 
 such components 
coherently interfere.
Furthermore, in Ref. \cite{noij2}~,
in order to show how much these conclusions  do not 
depend on the  model used, a semi factorized ansatz has been adopted in order 
to 
include non perturbative sea quarks and gluons at a given 
initial scale.  
For gluon-gluon dPDFs,  partonic correlations are still present also in 
the low-$x$ region. The strength of the correlations seems 
to become smaller but they 
are 
still sizable and should not be neglected.

\section{Calculation of  $\sigma_{eff}$ by means of CQM}

\begin{figure}[t]
\begin{center}
\epsfig{file=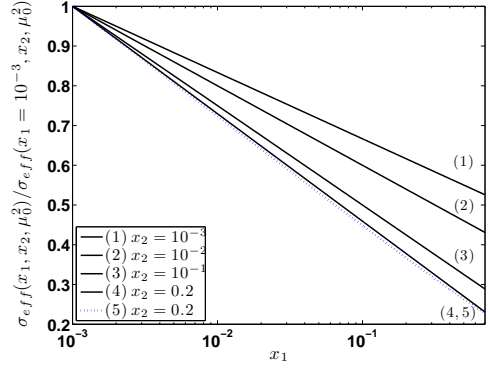, 
width=7.cm
}
\end{center}
\caption{ The ratio $
{\sigma_{eff}(x_1,x_2,\mu_0^2)}/{\sigma_{eff}(x_1= 
10^{-3},x_2,\mu_0^2)   }$    as a function 
of $x_1$ at fixed $x_2 = 0.001, 0.01, 0.1, 0.2.$}
\label{fff}
\end{figure}

A fundamental 
 quantity, relevant for the experimental 
analyses of DPS is the so called
 effective cross section, $\sigma_{eff}$ whose formal
 expression 
in terms of PDFs and dPDFs,
has been described in Ref. \cite{noiPLB}~.
Here,
$\sigma_{eff}$  has been evaluated  at 
 zero rapidity  through the dPDFs addressed in Ref. \cite{noij1} and calculated 
at $Q^2 = 250$ GeV$^2$ for different 
degrees of freedoms. 
The range of obtained values of $\sigma_{eff}$  is  comparable with 
the three old experimental extractions of
$\sigma_{eff}$ 
\cite{afs,data0,data2}~. Moreover, 
 as discussed in 
Ref. \cite{noice}~, if Melosh rotations were neglected, the average value of 
$\sigma_{eff}$ at the hadronic scale would change by a factor 2, making the 
impact of relativistic 
effects in $\sigma_{eff}$ very important. The main important result found and 
discussed in Refs. \cite{noiPLB}, is the  
strong $x_i$ dependence of 
$\sigma_{eff}$ in the valence region, a clear sign of DPC. In  Ref. 
\cite{noiads}~, $\sigma_{eff}$ has been studied also 
through dPDFs calculated within the AdS/QCD correspondence 
\cite{ads1,ads2,ads3}~. As one can see, the mean value 
of 
$\sigma_{eff}$ and its 
 $x$ dependence, obtained within this different model,  
 is comparable with those found within the LF 
approach in Ref. \cite{noiPLB}~. In order to emphasize this $x$ dependence, 
in 
Fig. 
\ref{fff}, the ratio $
{\sigma_{eff}(x_1,x_2,\mu_0^2)}/{\sigma_{eff}(x_1= 
10^{-3},x_2,\mu_0^2)   }$, evaluated by means of the AdS/QCD correspondence, 
has been plotted.
One can see that the ratio  strongly depends on 
$x_2$ in the valence region and longitudinal 
correlations cannot be neglected.

\section{Parton correlations in same sign $W$ pair production}
\begin{figure}[t]
\begin{center}
\epsfig{file=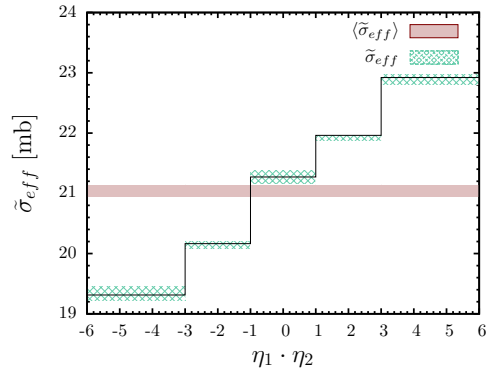, 
width=6.3cm
}
\end{center}
\caption{ $\widetilde{\sigma}_{eff}$ and $\langle 
\widetilde{\sigma}_{eff} \rangle$ as a 
function of product of muon rapidities. The error band represents scale 
variations added in quadrature.}
\label{efb}
\end{figure}

In this last section the main outcomes discussed in Ref. \cite{noiprl} will be 
summarised.
Due to results described in the previous section,
 we have studied to what extent double parton 
correlations could be accessed  at the LHC. To this aim we have considered the 
same sign $W$ pair production, a promising 
process for the observation of DPS \cite{kulesza, maina, gaunt2, campbell}~. 
The differential DPS cross section can be written as follows \cite{1a}~:

\begin{align}
 d \sigma^{AB}_{DPS} = \dfrac{m}{2} \sum_{i,j,k,l} d \vec z_{\perp} 
F_{ij}(x_1,x_2, \vec z_\perp, \mu)F_{ij}(x_3,x_4, \vec z_\perp, \mu)
&d\hat \sigma^A_{ik} d\hat \sigma^B_{jl}.
\label{dps}
\end{align}
In order to properly evaluate Eq. (\ref{dps}) by means of the use of dPDFs 
calculated in Ref. \cite{noij1}~,  it is worth to notice that in this scenario 
the  initial hadronic scale lies in the infrared region so  that the DPS cross 
section is extremely sensitive to its value.  Therefore, in Ref. 
\cite{noiprl}~, it 
has been  proposed fo fix the initial scale of the model ($Q_0^2$), used to 
calculate dPDFs appearing in eq. (\ref{dps}) as follows:
$i)$ we evaluated the 
single parton scattering cross section for $W$ production by using for PDFs the 
same LF CQM used for dPDFs of Ref. \cite{noij1}, $ii)$ then results found  are
compared  with 
those obtained by means of DYNNLO code\cite{catani} at LO within the PDF 
parametrization of Ref. \cite{MSTW}.
The best match is obtained for $0.24< Q^2_0<0.28$ GeV$^2$. We therefore choose 
  $Q_0^2=0.26$ GeV$^2$ as central value.
 Furthermore, uncertainties due to higher order corrections have been taken 
into account into the scheme by  also varying the final scale value, i.e., $0.5 
M_W<\mu_F<2.0M_W$, being $M_W$ the $W$ mass. 
 Details of the fiducial DPS phase space 
adopted in the analysis are discussed in Ref. \cite{noiprl}~. In particular we 
found that  the total $W$-charge summed DPS cross section (considering both 
$W'$s decays into same sign muons), calculated by means 
of dPDFs \cite{noij1}~, is found to be $\sigma^{++}+\sigma^{--} [\mbox{fb}] = 
0.69 \pm 0.18 (\delta \mu_F)^{~+0.12}_{~-0.16}(\delta  Q_0^2)$. This
result is consistent with those obtained by using for dPDFs the ansatz Eq. 
(\ref{app}) with PDFs evaluated with the MSTW parametrization 
\cite{MSTW} and those obtained with dPDFs of the model described in Ref. 
\cite{3a}~.  Further results of this analysis are presented in terms of the 
muon rapidities, i.e., $\eta_1 \cdot \eta_2 \simeq 1/4~ 
\mbox{ln}(x_1/x_3)\mbox{ln}(x_2/x_4)$. In particular, the DPS cross section 
converted in per-bin number of events, for a luminosity $\mathcal{L} = 300 
~\mbox{fb}^{-1}$, has a maximum for $\eta_1 \cdot \eta_2 \sim 0$ where 
annihilating partons share the momentum fraction in  at least one scattering. As 
one can see in Ref. \cite{noiprl}~,  results obtained within the LF CQM and with 
the prescription of Ref. \cite{3a}~, are compatible within the errors.
Moreover, the effects of DPC have been investigated
by observing the
 $\sigma_{eff}$ dependence on   $\eta_1 \cdot \eta_2 $. Hereafter we call 
such quantity, evaluated within the LF CQM, $\widetilde \sigma_{eff}$. In 
particular, for this process we found a mean value  $\langle \widetilde 
\sigma_{eff} \rangle = 21.04 \pm 0.07 (\delta Q^2_0)^{~+0.06}_{~-0.07}(\delta  
\mu_F)$ mb, consistently with calculations discussed in previous section. 
However, the most interesting  result is shown  in Fig. \ref{efb}. Here, 
in fact, we appreciate a clear signature of the presence of double 
correlations by observing the
departure of   $\widetilde \sigma_{eff}$ from a constant.  We therefore 
demonstrated that the experimental analysis of $\sigma_{eff}$ in bins of $\eta_1 
\cdot \eta_2$ is a suitable choice for studies on DPC. We have also estimated 
that, with a luminosity of $\mathcal{L} \sim 1000 
~\mbox{fb}^{-1}$, at 68\% confidence level, such a departure of $\sigma_{eff}$ 
from a constant value can be measured in the next run of the LHC.

\section{Conclusions}
In this contribution we have shown new results from the calculations of dPDFs 
evaluated within  a LF relativistic quark model. In particular, by using pQCD 
evolution of dPDFs, we found that longitudinal correlations affect dPDFs  
at very high momentum transfer  and in the small $x$ region, also for 
gluon-gluon 
distributions. Moreover, due to the correct treatment of the relativity, thanks 
to the Light-Front approach, also correlations between $x$ and transverse 
position $\vec z_\perp$ 
are found to be important and largely model independent.  The study of 
correlation 
effects in  $\sigma_{eff}$ shows that in the valence region  such a 
quantity is not constant and depends on $x$ and also using  the AdS/QCD 
correspondence for the calculation of 
dPDFs the result is confirmed. In order  to provide a detailed analysis 
of $\sigma_{eff}$ in a given process, we studied the same sign $W$ pair 
production process by calculating its cross section and $\sigma_{eff}$. 
Results we found are consistent with previous investigations and indicate that 
in this specific channel, correlations in dPDFs can be observed in the next run 
of the LHC. 
 This work was 
supported in part by the Mineco under contract FPA2013-47443-C2-1-P and 
SEV-2014-0398. This work is also supported though the project “Hadron Physics 
at the LHC: looking for signatures of
multiple parton interactions and quark gluon plasma formation
(Gossip project)”, funded by the “Fondo ricerca
di base di Ateneo” of the Perugia University.
The author thanks the organizers of the conference for the support given for 
this talk.

\end{document}